# High-temperature superconductivity in $Nd_{0.85}Sr_{0.15}NiO_2$ membranes under pressure

Yonghun Lee[1,2,§], Mengnan Wang[3,§], Xin Wei[1,4,§], Yijun Yu[1,2,5], Wendy L. Mao[1,3], Yu Lin[1,*], and Harold Y. Hwang[1,2,*]

[1]Stanford Institute for Materials and Energy Sciences, SLAC National Accelerator Laboratory, Menlo Park, CA 94025, USA

[2]Department of Applied Physics, Stanford University, Stanford, CA 94305, USA

[3]Department of Earth and Planetary Sciences, Stanford University, Stanford, CA 94305, USA

[4]Department of Physics, Stanford University, Stanford, CA 94305, USA

[5]Present address: State Key Laboratory of Surface Physics and Department of Physics, Fudan University, Shanghai, China

*Email: lyforest@stanford.edu (Y.Lin), hyhwang@stanford.edu (H.Y.H.)
[§]Y.L., M.W., and X.W. contributed equally to this work.

## Abstract

Lattice compression has emerged as a fundamental tuning parameter for nickelate superconductivity. Pressure acts as a trigger to induce superconductivity in bulk Ruddlesden-Popper nickelates[1,2]. For infinite-layer nickelate thin films[3], compressive epitaxial strain[4,5] and rare-earth ion chemical pressure[6,7] have been used to substantially enhance the superconducting transition temperature ($T_c$). Efforts to go further have been constrained by the limits of epitaxial stability or the challenges of measuring thin films in high-pressure environments[8]. Here, we overcome this limitation by developing a technique to incorporate freestanding infinite-layer $Nd_{0.85}Sr_{0.15}NiO_2$ membranes[9] into a diamond anvil cell. Using this platform, we observe a strong increase in $T_c$ up to our highest measurement pressure of ~90 GPa, where a superconducting downturn can be observed near liquid nitrogen temperatures. Strikingly, we find a simple linear enhancement of $T_c$ at a rate of 0.65 K GPa$^{-1}$, with no signs of saturation. This suggests that the pairing strength in infinite-layer nickelates can be raised to a surprisingly high scale, using an approach that can be broadly applied to many two-dimensional materials.

**Introduction**

Since the discovery of superconductivity in infinite-layer nickelates[3], considerable evidence has accumulated to indicate its unconventional nature, including the observation of a superconducting dome[4,10], the presence of strong magnetic fluctuations[11,12], and evidence for a nodal gap structure[13–15]. Already the initially reported superconducting transition temperature ($T_c$) was far above the scale expected from electron-phonon coupling[16]. Materials optimization in epitaxial thin films has progressively increased $T_c$ from the initial ~15 K (Ref. [3]) to ~40 K (Ref. [6]). Two developments have been central to this progress. First, moving to substrates better lattice-matched to the precursor perovskite phase improved crystallinity while imposing compressive strain on the infinite-layer phase[4,5,17]. Second, replacing the rare-earth ion with samarium better accommodated this smaller substrate lattice while introducing chemical pressure that shortened the $c$-axis lattice parameter[6,7]. Together, these advances establish a clear empirical trend – lattice contraction drives $T_c$ enhancement. This raises a fundamental question: what is the intrinsic upper bound of $T_c$ in infinite-layer nickelates, and does lattice-contraction-driven enhancement continue beyond the range currently accessible in thin films?

High pressure provides a natural route to address this question. The structural simplicity of the infinite-layer phase circumvents the complex structural responses associated with planar buckling in cuprates[18,19] and octahedral tilts in bilayer nickelates[1]. In this sense, it offers a cleaner platform to investigate how the electronic structure and superconductivity evolve under compression. Progress in high-pressure studies has been constrained, however, by the absence of bulk superconducting infinite-layer nickelate samples[20]. While a seminal thin-film pressure study reported an increase in $T_c$ up to ~30 K, the stability of these substrate-bound films rapidly deteriorated well below ~12 GPa (ref. [8]). This limitation motivates an alternative geometry: a freestanding membrane in which the superconducting infinite-layer nickelate film is fully released from its growth substrate[9,21]. This sample geometry is similar to exfoliated flakes from a broad family of two-dimensional materials, for which high pressure studies are beginning to emerge[22]. Here we use such membranes in a diamond anvil cell (DAC) that enables us to access the extreme pressure regime.

**Main**

Optimally doped $Nd_{0.85}Sr_{0.15}NiO_2$ thin films (~6.7 nm) were synthesized by pulsed laser deposition followed by topotactic reduction. The films were encapsulated by protective $SrTiO_3$ capping layers and grown on a water-soluble sacrificial layer (synthesis and characterization described in Ref. [9]). Dissolution of the sacrificial layer released a polymer-supported infinite-layer nickelate membrane, enabling direct integration of the superconducting membrane onto the diamond culet in a substrate-free geometry (Fig. 1a,b). Electrical contacts were defined in a van der Pauw geometry using a patterned, suspended silicon nitride stencil mask. Sequential ion milling and electron-beam evaporation locally removed the insulating $SrTiO_3$ cap to form Ti/Au Ohmic contacts to the membrane (Fig. 1c,d). The electrodes were extended using Pt foils, and the DAC was assembled with a metal gasket insulated by cubic boron nitride (cBN) powder (Fig. 1e) and silicone oil as the pressure-transmitting medium (PTM). The substrate-free membranes exhibit robust superconductivity, with $T_c$ rising continuously throughout stepwise pressurization. This pressure-driven enhancement culminates at ~91.5 GPa (our maximum achieved measurement pressure), where the temperature-dependent sheet resistance $R(T)$ shows a superconducting transition with $T_c \sim 74.2$ K (Fig. 1f; $T_c$ is taken as the onset temperature $T_{c,onset}$ throughout this work including literature comparisons, see Extended Data Fig. 1).

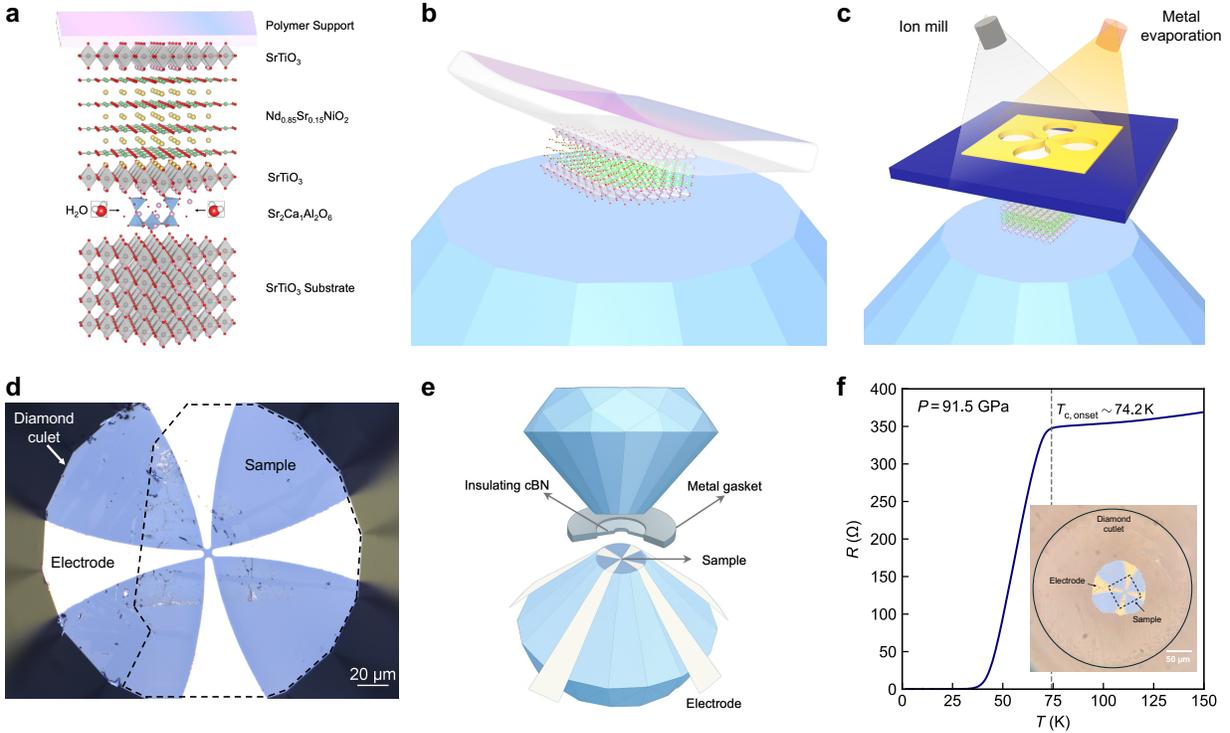

**Fig. 1 | High-pressure DAC platform for freestanding infinite-layer nickelate membranes. a**, Schematic of the polymer-supported 10 u.c. SrTiO$_3$ / 20 u.c. Nd$_{0.85}$Sr$_{0.15}$NiO$_2$ / 10 u.c. SrTiO$_3$ / 10 u.c. Sr$_2$Ca$_1$Al$_2$O$_6$ heterostructure grown on the SrTiO$_3$ substrate, followed by water etching of the sacrificial layer. **b**, Transfer of the freestanding membrane onto a diamond culet. **c**, Argon ion milling and electron-beam evaporation of Ti (5 nm)/Au (125 nm) onto the transferred membrane using a silicon nitride stencil mask. **d**, Optical microscope image of the processed sample on the diamond culet. **e**, Schematic of the DAC assembly with a metal gasket coated with insulating cBN. Pt foil extends the evaporated metal electrodes. **f**, $R(T)$ at $P = 91.5$ GPa, showing $T_{c,onset} \sim 74.2$ K. Inset: an optical microscope image of the sample chamber within the DAC.

Figure 2a,b displays the stepwise evolution of $R(T)$ with pressure for two representative samples measured using 300 μm and 200 μm diamond culets, respectively; the 200 μm culet was used to access higher pressures up to its ~90 GPa mechanical limit. At ambient pressure, $T_c \sim 17$ K falls in the established range reported for as-grown thin films and freestanding membranes transferred onto Si/SiO$_2$ substrates[9]. With increasing pressure, $T_c$ rises monotonically, a trend observed across all six measured samples (Extended Data Fig. 2). The zero-resistance temperature $T_{c,0}$ (defined as the temperature at which the resistance reaches the measurement noise floor) lags

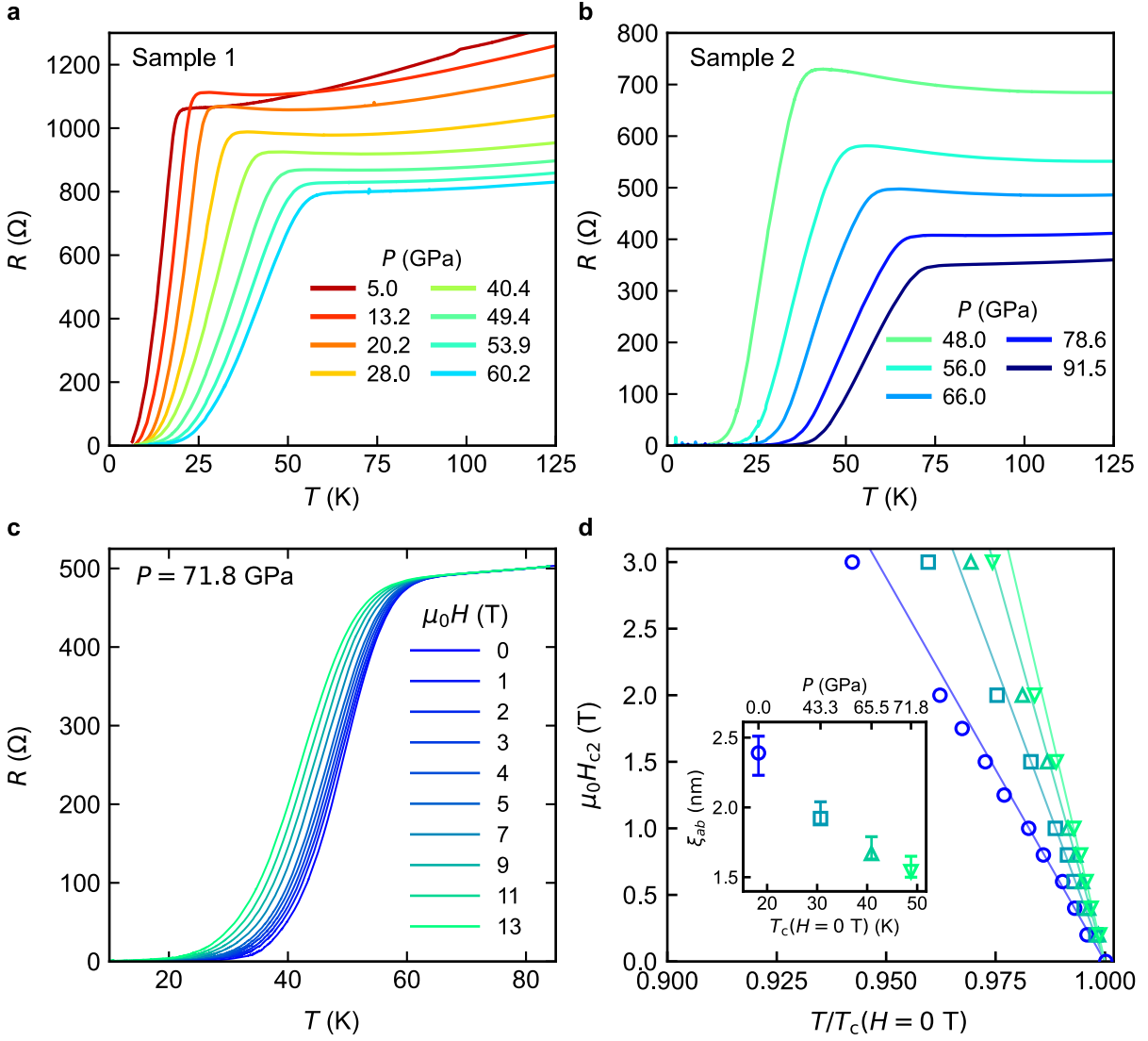

**Fig. 2 | Pressure evolution of superconductivity in infinite-layer nickelate membranes. a**, **b**, $R(T)$ as a function of pressure for Sample 1 on a 300 μm culet (**a**) and Sample 2 on a 200 μm culet (**b**). **c**, $R(T)$ of Sample 3, measured at 71.8 GPa under out-of-plane magnetic fields from 0 to 13 T. **d**, Upper critical fields $\mu_0 H_{c2}$ as a function of normalized temperature $T/T_c(H=0\text{ T})$ for a pristine thin film ($P=0$ GPa, circles) and Sample 3 at $P=43.3$ GPa (squares), 65.5 GPa (upward triangles), and 71.8 GPa (downward triangles). $H_{c2}$ values were determined using 50% of the extrapolated normal-state resistance criterion ($H_{c2}^{50\%}$). Solid lines represent linearized Ginzburg-Landau fits near $T_c$. The inset shows the extracted in-plane coherence length $\xi_{ab}$ plotted against zero-field $T_c$ (bottom axis) and the corresponding pressure (top axis).

the pressure evolution of $T_{c,\text{onset}}$, leading to a broadened superconducting transition. Such broadening is also observed in high-pressure transport measurements across cuprates[23] and Ruddlesden-Popper nickelates[1,24]. A likely origin is the non-hydrostatic environment in the DAC, arising from the solidification of the silicone oil PTM near its glass transition at a relatively low pressure (~3 GPa)[25]. The observed broadening cannot be explained by the small macroscopic pressure gradient alone (Extended Data Fig. 3), but may reflect the nanometer-thick membrane geometry, which has limited ability to elastically average over localized load variations.

Despite this expected sensitivity to non-hydrostaticity, the superconducting transition remains robust up to the highest pressure reached. Notably, $T_{c,0}$ begins to increase more rapidly above ~60 GPa and narrows the gap to $T_{c,\text{onset}}$, leading to a stabilization of the transition width (Extended Data Fig. 4). Ultimately, the accessible pressure range is limited by diamond anvil fracture rather than an apparent saturation of the sample response. The concurrent accelerated rise in $T_{c,0}$ and reduction in normal-state resistance suggests that inhomogeneities becomes less limiting at elevated pressures.

Figure 2c shows a series of $R(T)$ curves measured under increasing out-of-plane magnetic fields up to 13 T at $P = 71.8$ GPa, where $T_c$ reaches ~64 K. Below $T_c$, the magnetoresistive transition is progressively suppressed and broadened, characteristic of type-II superconductors. From these curves, we extract the upper critical field, $H_{c2,\perp}$, using the criterion of 50% of the extrapolated normal-state resistance (following the approach in Ref. [26]), and plot the values as a function of normalized temperature in Fig. 2d. From the $T$-linear response in the critical regime near $T_c$, we fit the data with the linearized Ginzburg-Landau form to obtain the in-plane coherence length, $\xi_{ab}$. Repeating these measurements at intermediate pressures, alongside a pristine thin film as a zero-pressure reference, reveals a clear change in the slope of the linear fits (Fig. 2d). The extracted $\xi_{ab}$ decreases from ~2.39 nm to ~1.54 nm at 71.8 GPa, exhibiting a monotonic reduction that inversely tracks the increase of $T_c$ under pressure, reflecting the strengthening of the superconducting state.

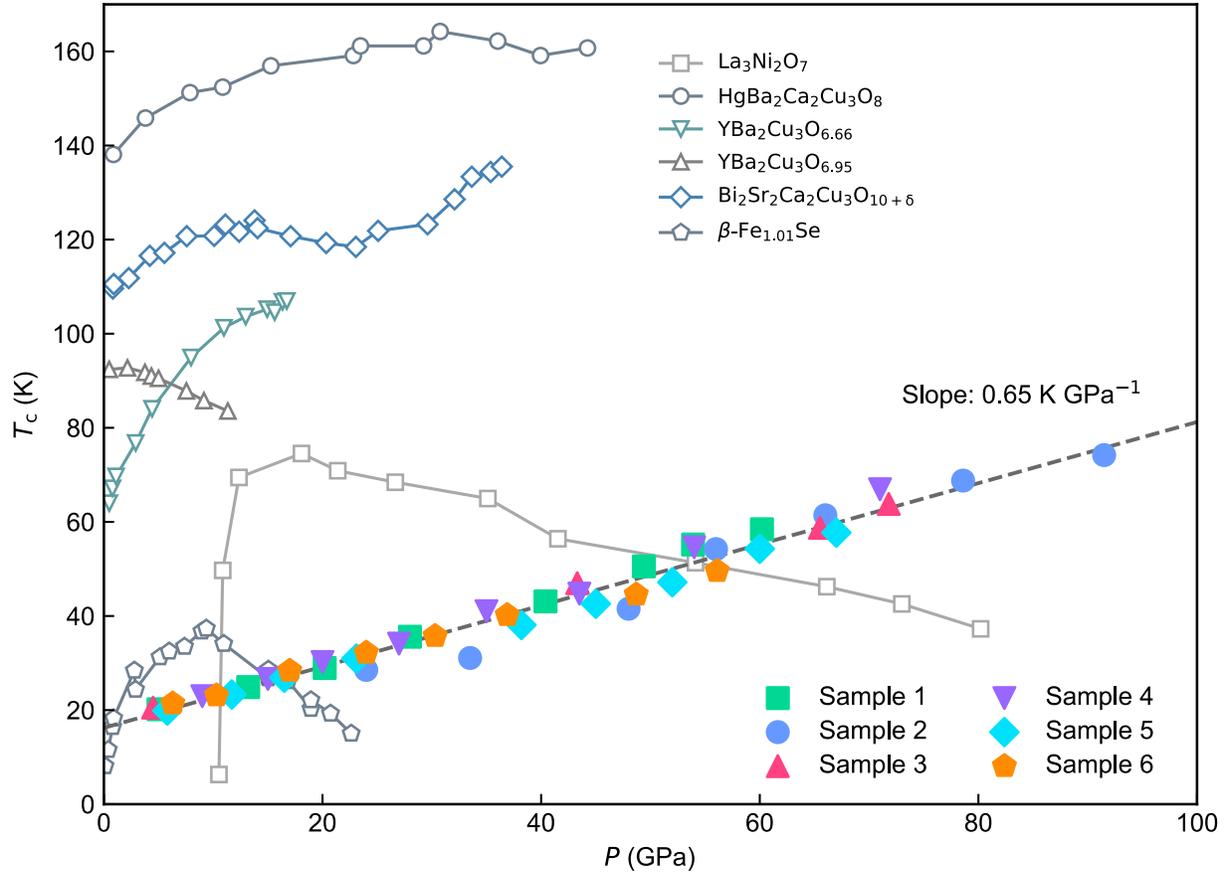

**Fig. 3 | Distinct high-pressure $T_c$ trajectory relative to other superconducting families.** Onset $T_c$ values of all six measured samples (closed symbols) are fitted together, yielding a combined linear increase of ~0.65 K GPa$^{-1}$ (dashed gray line). Open symbols provide comparison data for bilayer nickelate, various cuprates, and iron selenide: $La_3Ni_2O_7$ (ref. [24]), $HgBa_2Ca_2Cu_3O_8$ (ref. [23]), $YBa_2Cu_3O_{6.66}$ (ref. [27]), $YBa_2Cu_3O_{6.95}$ (ref. [27]), $Bi_2Sr_2Ca_2Cu_3O_{10+\delta}$ (ref. [33]), and $\beta$-$Fe_{1.01}Se$ (ref. [28]).

**Discussion**

Figure 3 compiles the pressure dependence $T_c(P)$ for all six measured samples, which are remarkably consistent. The data are well described by a line with slope $dT_c/dP \sim 0.65$ K GPa$^{-1}$. Surprisingly, $T_c$ exhibits no sign of saturation. This very simple linear increase motivates a comparison with the high-pressure response of other superconducting nickelates, cuprates, and iron selenides. Most high-pressure studies of high-$T_c$ superconductors report an eventual suppression of $T_c$ at elevated pressure. The bilayer nickelate $La_3Ni_2O_7$, for example, exhibits a sharp emergence of superconductivity near ~10 GPa followed by a gradual suppression beyond ~20 GPa (ref. [1,24]). Cuprates such as $HgBa_2Ca_2Cu_3O_8$ (Ref. [23]) and $YBa_2Cu_3O_{6.66}$ (Ref. [27]) and $\beta$-$Fe_{1.01}Se$ (Ref. [28]) similarly exhibit an eventual saturation and downturn in $T_c(P)$ (Fig. 3).

In cuprates, the pressure evolution is widely understood through two competing effects: a non-doping structural contribution that enhances pairing, and a pressure-induced charge transfer that ultimately overdopes the system and suppresses superconductivity[29]. Depending on the balance of these effects, optimally doped cuprates may exhibit a transient initial rise[23,30] or an almost immediate overdoping-driven suppression (e.g., $YBa_2Cu_3O_{6.95}$ (Ref. [27]) in Fig. 3). Theoretical models for infinite-layer nickelates propose a similar fate: a pressure-enhanced self-doping effect driven by $c$-axis compression of rare-earth-derived three-dimensional electron pockets[31,32] is predicted to cause $T_c$ in optimally doped $Pr_{0.82}Sr_{0.18}NiO_2$ to peak near 50 K at ~50 GPa before declining due to self-doping into the overdoped regime[31]. Against this context, not only does the over four-fold $T_c$ enhancement we observe for optimally doped $Nd_{0.85}Sr_{0.15}NiO_2$ far exceed the maximum increase seen in cuprates (~1.7), but the persistent, non-saturating trajectory has yet to exhibit any sign of an overdoping penalty that apparently limits cuprates. We note a rare exception to the usual dome-shaped $T_c(P)$ is the $T_c$ resurgence in $Bi_2Sr_2Ca_2Cu_3O_{10+\delta}$ (ref. [33], Fig. 3), with similar trends found in monolayer and bilayer counterparts[34]. Although the underlying mechanism remains debated, this suggests that the non-saturating response observed in $Nd_{0.85}Sr_{0.15}NiO_2$ is not an isolated anomaly, and it implies further opportunities to enhance $T_c$.

**Methods**

*Diamond Anvil Cell Preparation*: High pressure was generated using a non-magnetic diamond anvil cell made of Be–Cu alloy, equipped with either 200 μm or 300 μm culet diamonds. Insulating gaskets were prepared from T301 stainless steel or Re (250 μm thickness) and pre-indented to ~25 GPa. The full culet area was then laser-drilled, and a mixture of cBN powder and epoxy (1:10 by weight)[35] was packed into the indentation and compressed between the anvils to form a dense insulating layer. For 200 μm culets, the insulating layer was subsequently compressed to a thickness of ~25 μm and a sample chamber of ~80 μm diameter was drilled using an in-house laser-drilling system. For 300 μm culets, the insulating layer thickness was ~30 μm with the sample chamber diameter of ~120 μm. Following sample transfer and electrode fabrication (Fig. 1b-d), Pt foils and Cu wires were attached to extend the evaporated electrodes. The gasket was then realigned on the anvil, and the diamond anvil cell was assembled with an initial anvil separation of ~25–30 μm, estimated from optical interference fringes. Silicone oil served as the pressure-transmitting medium to maintain quasi-hydrostatic conditions.

*Pressure calibration*: Pressure was calibrated at room temperature using ruby fluorescence (for 0.1–50 GPa) and the diamond Raman edge (for 50–100 GPa). The uncertainty is ~0.1 GPa for ruby fluorescence and, for the diamond Raman edge, increases with pressure but remains below 5% up to the highest pressure reached in this study. Pressures above 50 GPa were measured at positions adjacent to the sample rather than directly on the sample (see Extended Data Fig. 3). Ruby fluorescence and diamond Raman spectra were collected using a Renishaw inVia Raman microscope with a spectral resolution of 2–4 cm$^{-1}$ and 514.5/633 nm excitation lasers at a laser power of ~5mW to minimize possible damage to the sample.

**Data availability**

All source data supporting the findings of this study are provided.

**Acknowledgements**

Y.L., M.W., and X.W. contributed equally to this work. The authors acknowledge Feng Ke for his contribution to the initial stage of this work, and Bai Yang Wang, Kyuho Lee, Brian Moritz, Cheng Peng, Jiarui Li, and Florian Theuss for useful discussions. This work was supported by the U.S. Department of Energy, Office of Basic Energy Sciences, Division of Materials Sciences and Engineering under Contract No. DE-AC02-76SF00515. Development of device/electrical assembly was supported by the Kavli Foundation, Klaus Tschira Stiftung, and Kevin Wells.

**Author contributions**

Y.L. and X.W. synthesized freestanding infinite-layer nickelate membranes, transferred them onto diamond culets, and prepared electrical contacts. M.W. designed and assembled the diamond anvil cells for high-pressure resistance measurements. X.W. developed the dry-transfer method and stage for the precise integration of membranes onto diamond anvil cell. Y.L and X.W. carried out transport measurements with assistance from M.W. All authors contributed to the writing of the manuscript. Y.L., Y.Y., W.L.M., Y.Lin, and H.Y.H. supervised the project.

**Conflict of Interest**

The authors declare no conflict of interest.

**References**

1. Sun, H. *et al.* Signatures of superconductivity near 80 K in a nickelate under high pressure. *Nature* **621**, 493–498 (2023).

2. Zhu, Y. *et al.* Superconductivity in pressurized trilayer $La_4Ni_3O_{10-\delta}$ single crystals. *Nature* **631**, 531–536 (2024).

3. Li, D. *et al.* Superconductivity in an infinite-layer nickelate. *Nature* **572**, 624–627 (2019).

4. Lee, K. *et al.* Linear-in-temperature resistivity for optimally superconducting (Nd,Sr)$NiO_2$. *Nature* **619**, 288–292 (2023).

5. Ren, X. *et al.* Possible strain-induced enhancement of the superconducting onset transition temperature in infinite-layer nickelates. *Commun. Phys.* **6**, 1–8 (2023).

6. Chow, S. L. E., Luo, Z. & Ariando, A. Bulk superconductivity near 40 K in hole-doped $SmNiO_2$ at ambient pressure. *Nature* **642**, 58–63 (2025).

7. Yang, M. *et al.* Enhanced superconductivity and mixed-dimensional behaviour in infinite-layer samarium nickelate thin films. *Nat. Commun.* **17**, 2761 (2026).

8. Wang, N. N. *et al.* Pressure-induced monotonic enhancement of $T_c$ to over 30 K in superconducting $Pr_{0.82}Sr_{0.18}NiO_2$ thin films. *Nat. Commun.* **13**, 4367 (2022).

9. Lee, Y. *et al.* Synthesis of superconducting freestanding infinite-layer nickelate heterostructures on the millimetre scale. *Nat. Synth.* **4**, 573–581 (2025).

10. Zeng, S. *et al.* Phase diagram and superconducting dome of infinite-layer $Nd_{1-x}Sr_xNiO_2$ thin films. *Phys. Rev. Lett.* **125**, 147003 (2020).

11. Lu, H. *et al.* Magnetic excitations in infinite-layer nickelates. *Science* **373**, 213–216 (2021).

12. Krieger, G. *et al.* Charge and spin order dichotomy in $NdNiO_2$ driven by the capping layer. *Phys. Rev. Lett.* **129**, 027002 (2022).

13. Gu, Q. *et al.* Single particle tunneling spectrum of superconducting $Nd_{1-x}Sr_xNiO_2$ thin films. *Nat. Commun.* **11**, 6027 (2020).

14. Cheng, B. *et al.* Evidence for d-wave superconductivity of infinite-layer nickelates from low-energy electrodynamics. *Nat. Mater.* **23**, 775–781 (2024).

15. Harvey, S. P. *et al.* Evidence for nodal superconductivity in infinite-layer nickelates. *Proc. Natl Acad. Sci. USA* **122**, e2427243122 (2025).

16. Nomura, Y. *et al.* Formation of a two-dimensional single-component correlated electron system and band engineering in the nickelate superconductor $NdNiO_2$. *Phys. Rev. B* **100**, 205138 (2019).

17. Lee, K. *et al.* Effects of stoichiometry and epitaxial strain on the stabilization of infinite-layer nickelates. *APL Mater.* **13**, 101105 (2025).

18. Jorgensen, J. D. *et al.* Structural features that optimize high temperature superconductivity. in *Recent Developments in High Temperature Superconductivity* (eds Klamut, J., Veal, B. W., Dabrowski, B. M. & Klamut, P. W.) 1–15 (Springer, Berlin, Heidelberg, 1996).

19. Yamada, N. & Ido, M. Pressure effects on superconductivity and structural phase transitions in $La_{2-x}M_xCuO_4$ (M = Ba,Sr). *Physica C* **203**, 240–246 (1992).

20. Puphal, P. *et al.* Topotactic transformation of single crystals: from perovskite to infinite-layer nickelates. *Sci. Adv.* **7**, eabl8091 (2021).

21. Yan, S. *et al.* Superconductivity in freestanding infinite-layer nickelate membranes. *Adv. Mater.* **36**, 2402916 (2024).

22. Pimenta Martins, L. G. *et al.* High-pressure studies of atomically thin van der Waals materials. *Appl. Phys. Rev.* **10**, 011313 (2023).

23. Gao, L. *et al.* Superconductivity up to 164 K in $HgBa_2Ca_{m-1}Cu_mO_{2m+2+\delta}$ ($m$ = 1, 2, and 3) under quasihydrostatic pressures. *Phys. Rev. B* **50**, 4260–4263 (1994).

24. Li, J. *et al.* Identification of superconductivity in bilayer nickelate $La_3Ni_2O_7$ under high pressure up to 100 GPa. *Natl. Sci. Rev.* **12**, nwaf220 (2025).

25. Klotz, S., Chervin, J.-C., Munsch, P. & Le Marchand, G. Hydrostatic limits of 11 pressure transmitting media. *J. Phys. D: Appl. Phys.* **42**, 075413 (2009).

26. Wang, B. Y. *et al.* Isotropic Pauli-limited superconductivity in the infinite-layer nickelate $Nd_{0.775}Sr_{0.225}NiO_2$. *Nat. Phys.* **17**, 473–477 (2021).

27. Sadewasser, S., Schilling, J. S., Paulikas, A. P. & Veal, B. W. Pressure dependence of $T_c$ to 17 GPa with and without relaxation effects in superconducting $YBa_2Cu_3O_x$. *Phys. Rev. B* **61**, 741–749 (2000).

28. Medvedev, S. *et al.* Electronic and magnetic phase diagram of $\beta$-$Fe_{1.01}$Se with superconductivity at 36.7 K under pressure. *Nat. Mater.* **8**, 630–633 (2009).

29. Neumeier, J. J. & Zimmermann, H. A. Pressure dependence of the superconducting transition temperature of $YBa_2Cu_3O_7$ as a function of carrier concentration: A test for a simple charge-transfer model. *Phys. Rev. B* **47**, 8385–8388 (1993).

30. Chu, C. W. *et al.* Superconductivity above 150 K in $HgBa_2Ca_2Cu_3O_{8+\delta}$ at high pressures. *Nature* **365**, 323–325 (1993).

31. Di Cataldo, S., Worm, P., Tomczak, J. M., Si, L. & Held, K. Unconventional superconductivity without doping in infinite-layer nickelates under pressure. *Nat. Commun.* **15**, 3952 (2024).

32. Sharma, S., Jung, M.-C., LaBollita, H. & Botana, A. S. Pressure effects on the electronic structure and magnetic properties of infinite-layer nickelates. *Phys. Rev. B* **110**, 155156 (2024).

33. Chen, X.-J. *et al.* Enhancement of superconductivity by pressure-driven competition in electronic order. *Nature* **466**, 950–953 (2010).

34. Deng, L. *et al.* Higher superconducting transition temperature by breaking the universal pressure relation. *PNAS* **116**, 2004–2008 (2019).

35. Funamori, N. & Sato, T. A cubic boron nitride gasket for diamond-anvil experiments. *Rev. Sci. Instrum.* **79**, 053903 (2008).

**Extended Data Figures**

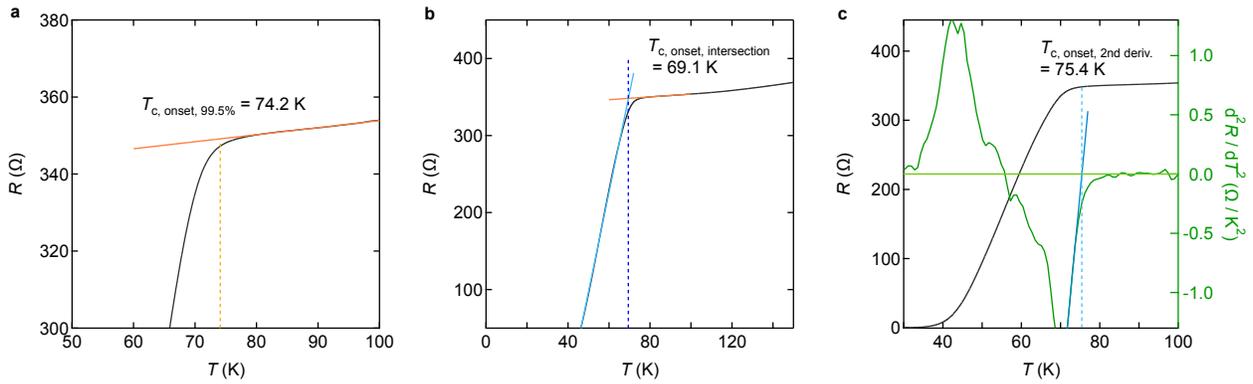

**Extended Data Fig. 1 | Evaluation method for the onset critical temperature $T_{c,\text{onset}}$. a**, The 99.5% criterion, defined as the temperature at which the resistance $R$ drops to 99.5% of the extrapolated normal-state linear fit. **b**, The intersection method, determined by the intersection of linear fits to the normal state and the superconducting transition region. **c**, The second-derivative method, defined as the zero-intercept of a linear fit to the $d^2R/dT^2$ curve near the onset region. Definition (**a**) is adopted throughout the manuscript.

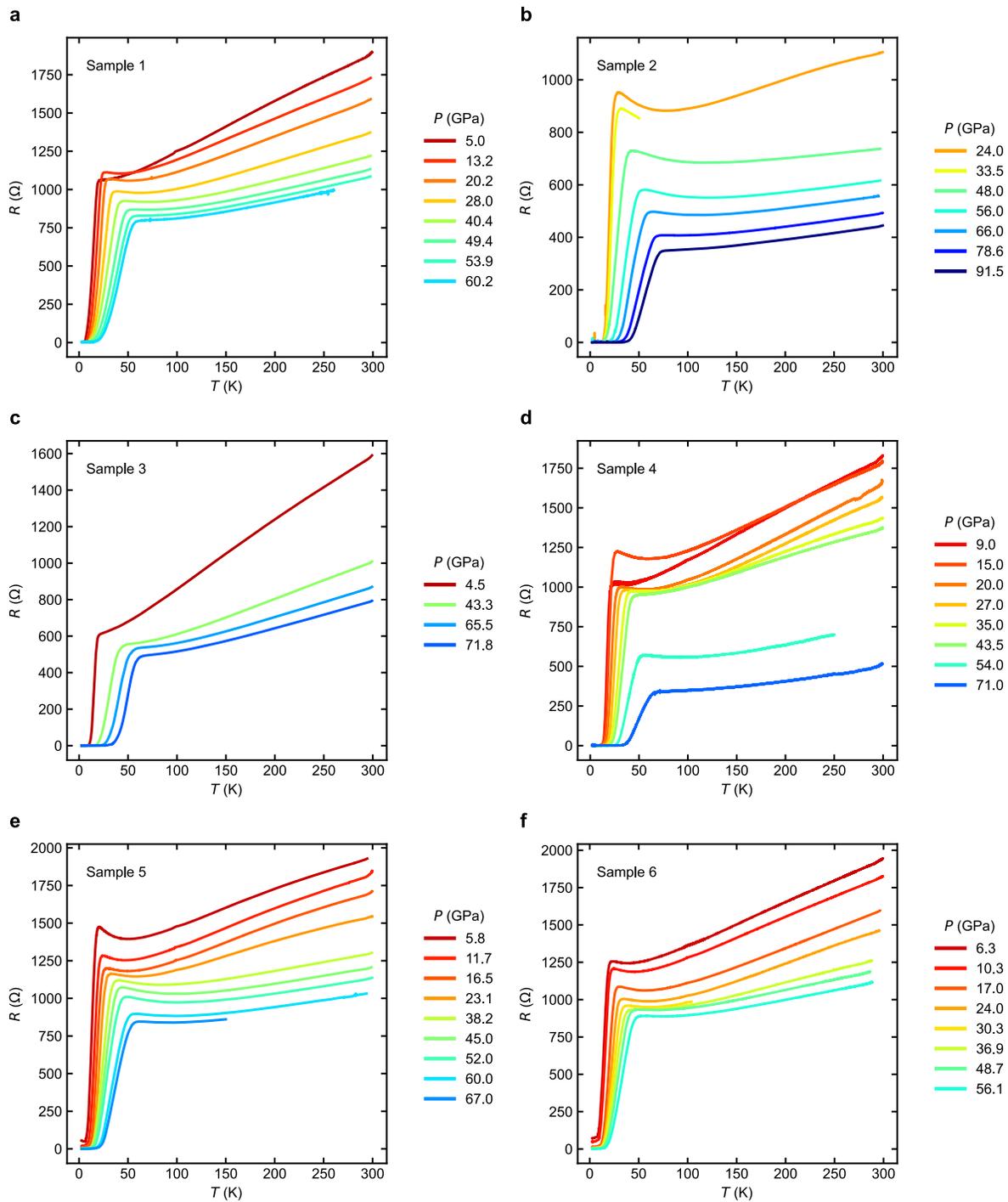

**Extended Data Fig. 2 | Full temperature range $R(T)$ for all samples. a-f**, sample 1–6, respectively.

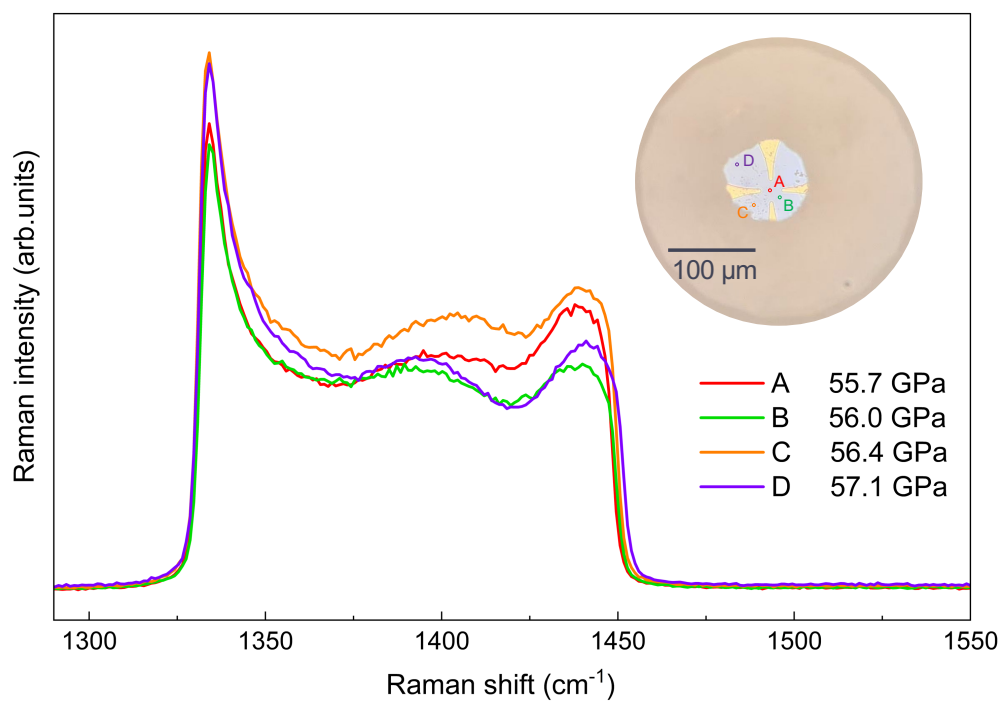

**Extended Data Fig. 3 | Pressure determination and pressure gradients in high-pressure electrical transport measurements.** Representative Raman spectra collected at ~56 GPa from different locations within the sample chamber are shown. A small pressure variation is observed across different locations. The reported pressure values represent the pressure in the immediate vicinity of the sample, estimated from Raman measurements taken at a nearby position (position B) to avoid direct laser exposure of the sample.

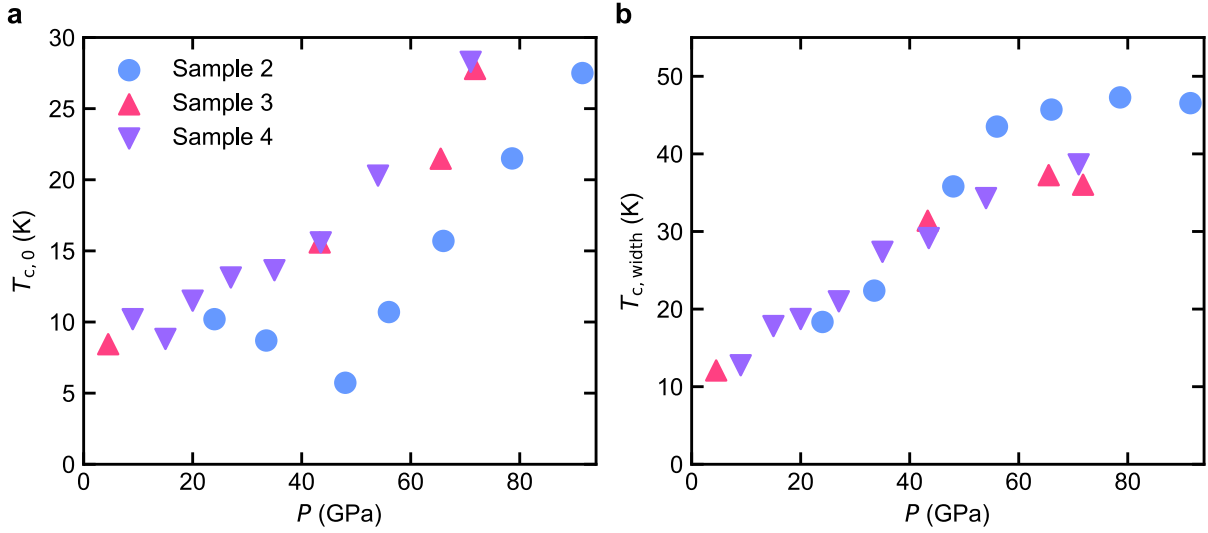

**Extended Data Fig. 4 | Enhancement in $T_{c,0}$ and corresponding saturation of the transition width. a**, $T_{c,0}$, defined as the temperature at which the resistance reaches the noise floor, as a function of pressure. Data are shown for Samples 2-4, where the zero-resistance state could be unambiguously resolved. **b**, Corresponding transition width, defined as $T_{c,\text{width}} = T_{c,\text{onset}} - T_{c,0}$.